# Persistent topological features of dynamical systems

Slobodan Maletić[1,2], Yi Zhao[1], Milan Rajković[2]

[1]Shenzhen Graduate School, Harbin Institute of Technology, Shenzhen, China

[2]Institute of Nuclear Sciences Vinča,

University of Belgrade, Belgrade, Serbia

Abstract

A general method for constructing simplicial complex from observed time series of dynamical systems based on the delay coordinate reconstruction procedure is presented. The obtained simplicia complex preserves all pertinent topological features of the reconstructed phase space and it may be analyzed from topological, combinatorial and algebraic aspects. In focus of this study is computation of homology of the invariant set of some well known dynamical systems which display chaotic behavior. Persistent homology of simplicial complex and its relationship with the embedding dimensions are examined by studying the lifetime of topological features and topological noise. The consistency of topological properties for different dynamic regimes and embedding dimensions is examined. The obtained results shed new light on the topological properties of the reconstructed phase space and open up new possibilities for application of advanced topological methods. The method presented here may be used as a generic method for constructing simplicial complex from a scalar time series which has a number of advantages compared to the the mapping of the same time series to a complex network.

The state-space reconstruction of a dynamical system from a scalar time series involves determination of the optimal delay time and the appropriate embedding dimension in order to ensure a diffeomorphic mapping between the true and the reconstructed attractor. Most of the important properties of the attractors of dynamical systems are topological and their topological structure may be revealed by constructing simplicial complex from the numerically or experimentally obtained time series. A method for the construction of a coarse-grained simplicial complex is presented which involves mapping of a certain number of closely distributed points contained in a high dimensional ball of a phase space to a single point, so that the procedure becomes less computationally demanding while preserving the intended topological properties. From the trajectories of several classic chaotic systems the dynamics is reconstructed using the method of delays and the coarse-grained Čech simplicial complex is constructed. Persistent homology is evaluated for several different embedding dimensions with the purpose of monitoring the lifetime of the basic topological features, the n-dimensional holes. The role of topological noise (short-lived topological structures) is examined and the observed features are interpreted from the aspect of the selection of appropriate reconstruction parameters.

## I. INTRODUCTION

Recent advances in computational topology methods [1] enable a number of possibilities for a novel approach to the problem of phase space reconstruction of dynamical systems from an observed time series and of characterization of various phase space properties. The reconstruction of phase space is of great importance for understanding complex dynamical systems as it indirectly involves several other directions of research. The extension of the standard delay reconstruction procedure presented here, being rooted in the geometrical and algebraic topological methods and concepts, sets a stage for new directions of research on the properties of dynamical systems. Although the approaches originating from geometrical and topological perspective are not novelty in phase space reconstruction [2], the recent advancements in the algebraic topology provide new tools for data analysis [1] and characterization of complex systems [3], and most importantly they are convenient for applications in the theory and analysis of nonlinear dynamic systems.

Unlike the case of mathematical models where the trajectory in phase space can be reconstructed directly from differential (or difference) equations, the reconstruction of phase space of real-world dynamical systems is not so simple and straightforward. The available information about the evolution of real-world nonlinear dynamical system is encoded in the experimental time series. Therefore, the challenge lies in the development of adequate and consistent methods for the reconstruction of phase space from time series, as well as for understanding its geometrical [2] and topological [4] structure, which would lead to better understanding, description and prediction of behavior of underlying complex dynamic system. The cornerstone and the starting point for development of such methods is the Takens embedding theorem [5] the proof of which indicates that the dynamics of smooth finite-dimensional dynamic systems evolving on smooth manifolds may be reproduced from the relevant time series of measurements performed on an evolving system. This method was later expanded in [6] to apply to a more general case when the dynamics evolves on fractal sets, with particular relevance for reconstruction of attractors in a finite-dimensional phase space. Since the proof of Takens theorem is based on the time delay embedding of the data in Euclidean space determination of the embedding dimension is closely related to the choice of the optimal delay time. Many different methods for phase space reconstruction have been developed although there is no consensus on the most favorable way to choose either the optimal time delay and/or the embedding dimension [7]. The more popular choices, for example, are the mutual information method for estimation of the optimal time delay [8], and false nearest neighbor method for the determination of proper embedding dimension [9].

In recent years the attention has been moved from solely reconstructing the qualitative features of the phase space to tendency for deeper understanding of geometrical and topological structure. The link that connected time series and the structure of phase space lies in another growing field of research, that is, the complex networks [10]. Namely, different methods have been developed for mapping a time series into a complex network and all of them proved to be useful for understanding of the dynamical system evolution. Depending on the way how the nodes and links are associated with the characteristics of dynamical system we mention here three of the more widely used networks: proximity networks, visibility graphs and transition networks [11]. Proximity networks are constructed from time series based on properties of reconstructed phase space, such as vectors in phase space or portions of time series. The representatives of proximity networks are the recurrence networks [12] and cyclic networks [13]. In the visibility graphs the stored information is related to the specific patterns emerging in time series [14]. Finally, the transition networks describe transitions between (meta)stable states of dynamical systems and are constructed from time series based on transition probabilities[15]. Since these approaches have been demonstrated as useful and practical in understanding complex dynamical systems, the attention has been turned toward mapping of time series into other types of complex networks, such as the weighted directed networks [16]. Transformations in the opposite direction have been explored as well, i.e., construction of time series from complex networks [17], and criteria based on geometrical invariability [18] have been introduced to confirm that the dynamical character of time series is preserved when mapped into a complex network.

Meanwhile, recognition of the importance and usefulness of simplicial complexes as convenient mathematical objects for scientific inquiry is slowly gaining ground. The importance of simplicial complexes and their topological and combinatorial properties in physics has been known for some time. Some of the prominent applications include the work of R. Atkin on applications in social systems [19] and discretization of physics [20], or the work of T. Regge on general relativity [21] and the applications in computational electromagnetism [22], to name a few. More recently, some of the applications include the alternative study of complex networks [3], topological data analysis [23], opinion dynamics [24] or topics in the study of social networks [25], [26]. One of the basic qualitative information that may be obtained by constructing a simplicial complex from a time series of a dynamical system is the dimension of the state space (the number of degrees of freedom of the system). Another basic information concerns the topology of the space. And although both of these basic properties are mostly qualitative, they are useful as, for example, in the case of a multifrequency periodic system. For such a system the geometry of the state space is an $n$-torus, where $n$ denotes the number of independent, possibly nonlinear, oscillators. Determination of the homology group would give information on the topological invariant characteristic of the system which may be used for recognition purposes. We extend these basic qualitative information by the application of persistent homology [27] and the quantities emerging from the Q-analysis [19], [28].

One way to construct a simplicial complex from a time series is first to map the time series into a complex network and then to construct a simplicial complex from the obtained network. The advantage of this method is that several different simplicial complexes may be constructed from a single network [3], [29], while the shortcomings are related to the time and computational requirements needed to perform the desired transformations. A more efficient approach which we pursue here is to construct a simplicial complex directly from the observe time series. More specifically, the construction includes a sequence of gradually nested simplicial complexes embedded in spaces of different dimensions determined by the embedding procedure.

It was during the course of writing of this paper that the authors became aware of two papers [30] and [31] with the similar subject matter. Although the focus of attention of these works includes the reconstruction of phase space based on the persistent homology of a specific type of simplicial complex, the so called witness complex [32], the purpose and the approach are essentially different from ours. The main interest of our work lies in the higher-order structures which emerge in the phase space topology expressed through preservation of the topological noise at different dimensions. The motivation of [30] is development of the method to avoid the full diffeomorphic reconstruction and to diminish the computational complexity, while of [31] is clustering of time series which have similar recurrent behavior. Nevertheless, the results of these two works and the present contribution are complementary and should bring to light advantages of the topological approach and perspectives for further advancements.

## II. THEORETICAL BACKGROUND

### A. Phase space reconstruction method

One of the important findings of dynamical systems theory is the existence of strange attractors in the phase spaces of nonlinear systems. Usually, the information about the dynamics is obtained from direct observations of relatively small number of dynamic variables, and the important contribution in this direction is that of "embedology" [6], a branch of dynamical systems theory which addresses an important problem of determining certain features of phase space solely from time series of a single observable. However, the realm of embedology is not a completed field of research in spite of reliable techniques which have been developed and successfully used for some time. In the standard procedure the starting point is an observed scalar time series, say $X = \{x_1, x_2, ..., x_n\}$ of some variable X measured at even time intervals and the signal is lifted to $d$ dimensions using time-delay embedding [5], so that the signal becomes a path in a higher-dimensional phase space of unknown dimension and shape. The points in the reconstructed space usually converge to a manifold or some other subspace of the $n$-dimensional Euclidean space or attractors of fractal dimension [6]. Note that Taken's theorem assumes idealized conditions in the sense that the observed time series is not contaminated by noise. From the practical aspect, the procedure of mapping the state vector of a dynamical system to a point in the reconstructed space consists of taking the $m$ uniformly spaced samples of the observed time series of the appropriate variable and concatenating them into a single vector:

$$\vec{v}_i(t) = [x_t, x_{t-\tau}, x_{t-2\tau}, ..., x_{t-(m-1)\tau}].$$

The key parameters in the reconstruction procedure are $\tau$, the time delay, and $m$, the embedding dimension. Taken's embedding theorem guarantees the preservation of topological properties of the attractor however not of its geometrical properties. This implies that the choice of the time delay and the embedding dimension have an important impact on the quality of the obtained results in applications. A requirement of the embedding theorem is that the embedding dimension should satisfy $m > 2d$, where $d$ is the real dimension of the dynamical system, that is, the true number of variables. The first step in the reconstruction procedure is the estimation of the (optimal) time delay $\tau$, and once this goal has been achieved the appropriate embedding dimension *m* is determined. The nontrivial aspect of phase space reconstruction is reflected in the large number of suggested techniques for estimations of τ and *m*. Two most often used methods for estimation of the delay time are based on determination of the first minimum of either the average mutual information [8] or autocorrelation function [33]. The advantage of the former is that both linear and nonlinear correlations are taken into account while the later includes only linear correlations. For the purposes of this study, the first local minimum of the mutual information is used to determine the optimal delay time.

The next step is the estimation of embedding dimension. For each value *m* = 2, 3, 4, ... of embedding dimension a simplicial complex is constructed which captures the topology of the reconstruction space followed by the calculation of topological invariants and properties using methods of the topological data analysis [1].

## B. Simplicial complex

Let $v_i \in R^m$ be the points in *m*-dimensional space. The *q*-simplex σ is defined as the convex hull of *q*+1 affinely independent points $\sigma = \langle v_0, v_1, \ldots, v_q \rangle$, and the points $v_i$ are called vertices [35]. Since the *q*-simplex σ builds a *q*-dimensional subspace of $R^m$ its dimension is *q* (dimσ = *q*). Any subset of a simplex, called a face, is also a simplex, since any subset of affinely independent points inherit that property. If we impose the ordering of vertices of a simplex (and accordingly the faces), an oriented simplex is defined. The geometric representation of a simplex is a polyhedron [35], for example, 0-simplex is a vertex, 1- simplex is an edge, 2-simplex is a triangle, 3-simplex is tetrahedra, etc. (Fig. 1).

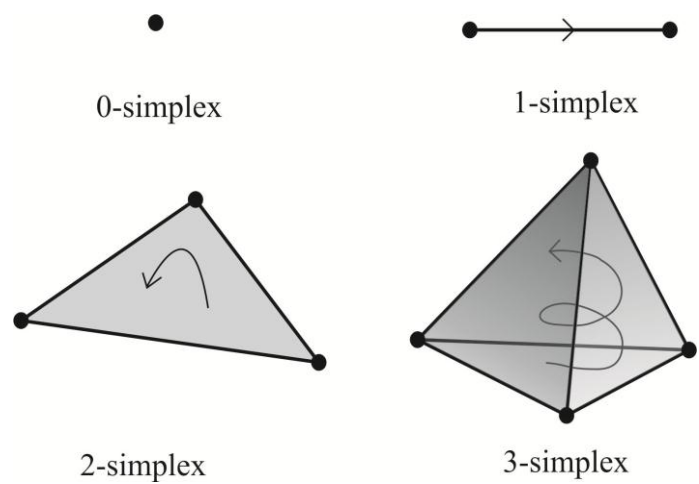

FIG. 1. Oriented 0-, 1-,2-, and 3-dimensional simplices.

The collection of simplices, together with their faces is called a simplicial complex, K, which is therefore closed under inclusion. In other words, simplicies, all of their faces and nonempty intersections of simplices (which are faces) are the building blocks of a simplicial complex. In the geometrical representation, simplicial complex is constructed by gluing polyhedra along the common faces. From this general definition it is obvious that in practice the rules for construction of a complex depend on the context and desired topological invariants and properties of the specific problem to be resolved. At this point it is useful to define an important topological invariant, the Betti number [35], which is helpful in distinguishing topological spaces.

Let $q$-chain be a collection of $q$-simplices in $K$, and a formal sum of two $q$-chains is also a $q$-chain. The collection of all $q$-chains together with the formal sum as a binary relation constructs a chain group $C_q$ [35] with properties of a vector space. If we define a boundary of a $q$-simplex $\sigma_q$ as a set of its $(q-1)$-faces (see Fig. 2), then the boundary of a $q$-chain is the sum of their $(q-1)$-faces, and hence, it is a $(q-1)$-chain. Formally, a $q$-th boundary operator is a mapping from $q$-chain group to $(q-1)$-chain group:

$$\partial_p : C_p \to C_{p-1}.$$

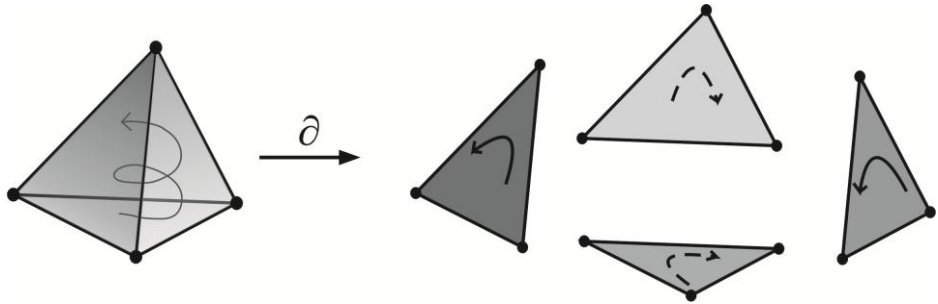

FIG. 2. Boundary operator maps a 3-simplex to four 2-simplices.

A gradual sequence of chain groups and boundary mappings between them is called the chain complex of $K$:

$$\emptyset \to C_q \xrightarrow{\partial_q} C_{q-1} \xrightarrow{\partial_{q-1}} \ldots \xrightarrow{\partial_2} C_1 \xrightarrow{\partial_1} C_0 \xrightarrow{c_0} \emptyset.$$

In order to define Betti numbers a definition of a homology group is required. Let $Z_q$ be the subgroup of $C_q$, such that its elements are all those $q$-chains whose boundary is empty. The elements of group $Z_q$ are $q$-cycles. Further, group $B_q$ is another subgroup of $C_q$, such that its elements are the boundary of a $(q+1)$-chain (i.e., $b \in B_q \subset C_q, c \in C_{q+1}, b = \partial_q c$). Finally, the $q$-th homology group $H_q$ is defined as: $H_q = Z_q/B_q$. A homology group is a collection of all those $q$-cycles that are not boundary of $(q+1)$- simplices. The rank (i.e., the number of homology group generators) of q-th homology group is called the $q$-th Betti number, $\beta_q$ = rank$H_q$. Less formaly, the $k$-th Betti number reveals the number of unconnected k-dimensional components. Intuitively, $\beta_0$ represents the number of connected components, $\beta_1$ represents the number one-dimensional (circular) holes, $\beta_2$ is the number of two-dimensional holes (voids), etc..

### C. Persistent homology method

The arbitrariness in simplical complex formation leaves ample freedom for choosing the appropriate construction procedure which could be adequate for various applications and topological requirements. A procedure of gradual formation of a simplicial complex starting from the simplest to the final one which is structurally stable is known as the filtration which is performed through the sequence of nested subcomplexes. Each subcomplex is contained in the next one of the filtration stage. Such method lies at the core of the computational algebraic topology approach called persistent homology [27]. More formaly, an increasing sequence

$$\emptyset = K_0 \subset K_1 \subset \cdots \subset K_n = K,$$

is the filtration of a simplicial complex $K$, and it can be understood as an addition of new simplices usually by changing some free parameter. Through the transitions from one filtration stage to another various topologically relevant changes occur, one of the most important being the creation/destruction of holes (i.e., increase/decrease of Betti numbers). Practitioners use different filtration procedures for building a simplicial complex [23] according to the context and the origin of the data. In the present study we have chosen a Čech complex filtration which is based on the change of radii of balls whose centers are associated to the points in the phase space. If two balls overlap then points in the phase space are connected by an edge, and if three balls overlap, they form a triangle (2-simplex), as illustrated in Fig. 3. In the same figure we show a situation when increasing the balls associated to the four points overlap to form a tetrahedron (3-simplex). The filtration of a given dataset is generated in the course of changing a single parameter, the ball radius $r$, and as a consequence homology classes are born (that is Betti numbers are increased), whereas some homology classes die (that is Betti numbers are decreased).

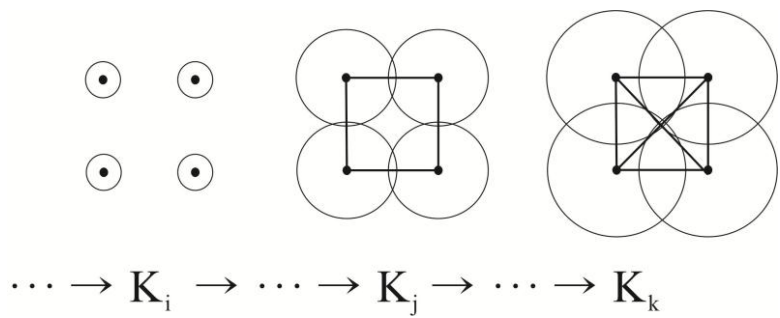

FIG. 3. Stages in building a Čech complex (tetrahedron) by increasing the radius over four points.

For visualization of persistence intervals of the homology group generator χ (that is, the interval from its birth in filtration stage $i_\chi$ to its death in filtration stage $j_\chi$) we use persistence barcodes [36] and persistence diagrams [27]. In persistence barcode the horizontal axes represents the change of filtration parameter, vertical axes represents homology group generators, and the persistence interval (or bar) is represented as a horizontal line associated to the *p*-th homology generator's birth/death filtration stage. Hence, for each *p* we have a sequence of horizontal lines with different lengths (Fig. 4). In the persistence diagram abscissa is associated with radius values of the birth of homology generators, and ordinate is associated with radius values of the death of homology generators. Hence, the coordinates of a point in persistence diagram represent birth and death of a particular generator (i.e., a *q*-dimensional hole) of the *q*-th homology group. Short bars in persistence barcode and points on and near diagonal in persistence diagram are associated with short-lived *q*-dimensional holes and they represent *topological noise*.

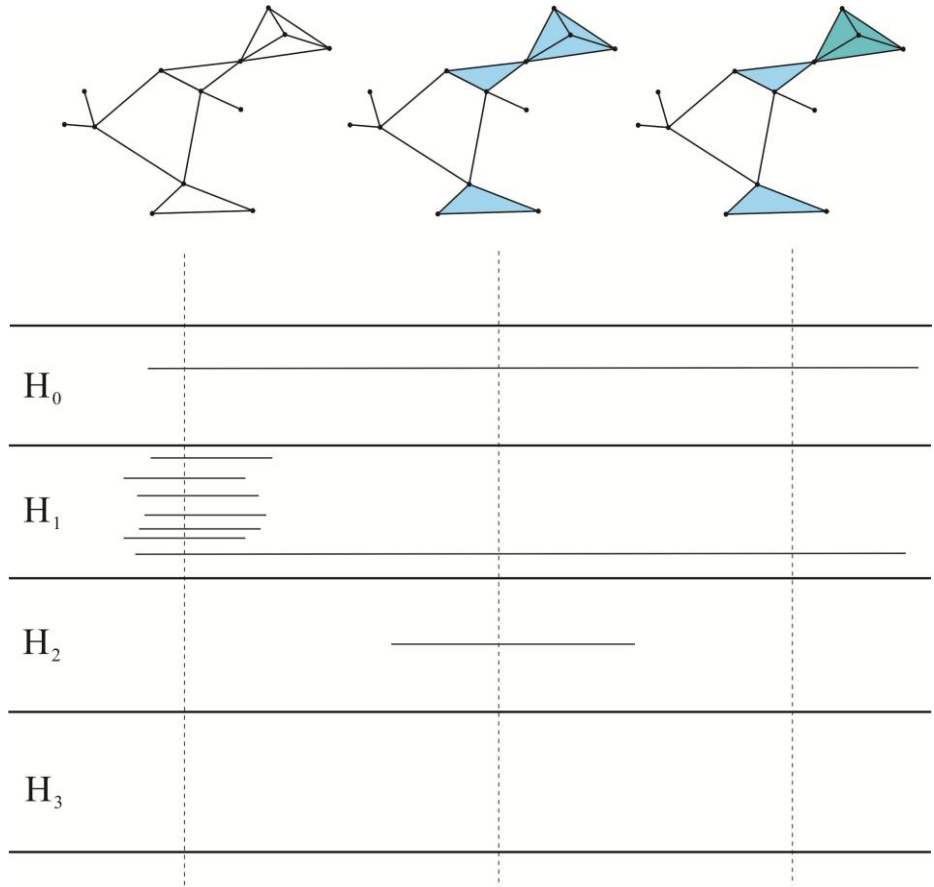

FIG. 4. An example of filtration and the barcode of an arbitrary simplicial complex. The intersection of the vertical dashed line and the horizontal line for different $H_k$ equals the rank of $H_k$, and persistent homologies are $H_0$ and $H_1$.

### D. Q-vector

An important quantity which characterizes intrinsic relational structure between collections of simplices is the *Q-vector* [19], whose statistical scaling features have been recently explored in the context of simplicial complexes [28]. Consider a chain of simplices so that each two consecutive ones share at least a *q*-face. Two simplices $\sigma_p^i$ and $\sigma_s^j$, having dimensions $p$ and $s$ respectively, are $q$-connected if between them is a sequence of $q$-face-shared simplices, which means that any pair of consecutive simplices share at least $q + 1$ vertices. The $q$-connectivity between simplices in a complex $K$ induces an equivalence relation which is reflexive, symmetric, and transitive. Let us denote this binary relation by $\epsilon_q$ and then

$$(\sigma^i, \sigma^j) \in \epsilon_q \text{ iff } \sigma^i \text{ is } q\text{-connected to } \sigma^j.$$

Let us denote by $K_q$ the subset of $K$ such that all simplices in $K_q$ have dimension larger than or equal to $q$. Applying the equivalence relation $\epsilon_q$ we partition $K_q$ into equivalence classes of $q$-connected simplices. The equivalence classes are then members of the quotient set $K_q/\epsilon_q$, and every simplex in one class is $q$-connected to every other simplex in that class without overlapping between classes. The cardinality of $K_q/\epsilon_q$, labeled by $Q_q$, enumerates the number of $q$-connectivity equivalence classes in a simplicial complex $K$. The value $Q_q$ is the $q$-th entry of Q-vector [19], an integer vector with the length $\dim(K) + 1$.

The Q-vector carries information on the structural relationship between simplices and we use it to make comparison between different topological structures emerging in different dynamical regimes. We assume that normalized Q-vectors are points in an abstract space, the Q-space, and we quantify similarity of two structures with the Euclidean distance between their corresponding Q-vectors.

### E. The search for topological signatures

Construction of simplicial complex from points in phase space can be extremely computationally demanding due to the emergence of large number of simplices and high simplicial dimensions, so some sort of coarse-graining is necessary. In order to overcome this obstacle we devised a coarse-graining procedure which consists of two steps:

(1) replacement of the group of 10 to 15 initial data points by a point at the center of the smallest Euclidean ball containing the initial set of points. The exact number of replaced points depends on their distribution and density in the phase space.

(2) application of Čech complex filtration on the new data set. A Čech complex is a simplicial complex formed such that a $k$-dimensional simplex is added any time a subset of $k$ points with common intersection of balls of certain predetermined radius formed around each point is encountered.

Note that the correlation sum [37] of the new and the coarse-grained data have the same scaling behavior for chaotic systems, due to the fractal nature of the strange attractors. The correlation sum for some set of $N$

points in *d*-dimensional space $\{\vec{x_1}, \vec{x_2}, \ldots, \vec{x_N}\}$ is defined by [37]

$$C_d(l) = \lim_{N \to \infty} \frac{1}{N^2} \sum_{i \neq j} \theta(l - \|\vec{x_i} - \vec{x_j}\|),$$

where θ(…) is the Heaviside function. When the data size is infinite and *l* small, the correlation sum scales like $C(l) \sim l^D$, where *D* is called correlation dimension. This scaling relationship remains the same when coarse-graining is performed since the condition requiring large number of points is fulfilled for the data under consideration, and due the fractal property of the systems used in the study.

The coarse-graining procedure may be interpreted as a special case of the reconstruction theorem [34]. Namely, the reconstruction theorem considers a set of neighboring points *M* on the attractor of a dynamical system, where neighborhood is defined as a distance smaller than a certain threshold value. If there is a mapping from this set *M* onto another set *N* such that two points are neighbors in *N* only if they are neighbors in *M*, then the mapping is homeomorphism and the two sets are topologically equivalent. In our calculations the neighboring distance between points is determined by the ball radius which is not fixed and depends on the distribution of points. An illustration of the procedure is presented in Fig. 5 for a dynamics of a simple pendulum in 2-dimensional embedding space. The initial data are represented as points and the new points are represented with asterisks. It is obvious that both sets of points preserve the topological features of the dynamics since both are distributed along a closed orbit (limit cycle).

The procedure of persistent homology through the Čech filtration is applied after coarse-graining, and topological noise as well as persistent invariants are detected. In the construction of the Čech complex the radius around each point is increased enlarging its immediate neighborhood. As the circles around two points overlap to a certain degree, holes appear like the islands of inaccessability (i.e., obstacles). By further expanding the radius holes may continue to exist or they disappear. The subspace of the phase space corresponding to the short-lived hole may be interpreted as temporary inaccessible (temporary obstacles), whereas persistent holes display a signature of permanent obstacles.

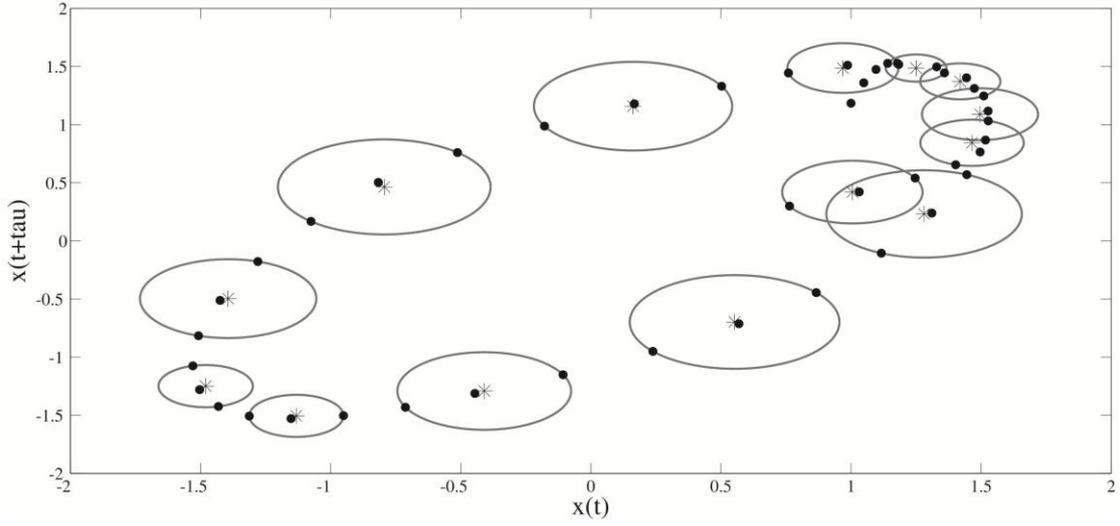

FIG. 5. The coarse-graining procedure for data points of a pendulum in a 2-dimensional embedding space. The initial points are represented as dots, whereas new points (i.e., centers of circles) are represented with asterisks.

## III. APPLICATIONS OF THE METHOD

In this section we illustrate the proposed method for the well known dynamical systems displaying chaotic behavior. Persistent homology is calculated using computational topology package "Plex" [38].

### A. Hénon map

The Hénon map [39] is a 2-dimensional iteration map given by equations

$$x_{i+1} = 1 + y_i - \alpha x_i^2$$

$$y_{i+1} = \beta x_i$$

(1)

where $\alpha > 0$ and $|\beta| < 1$. For particular values of parameters $\alpha = 1.4$ and $\beta = 0.3$ the system exhibits chaotic behavior with a characteristic shape of the strange attractor (Fig. 6). A time series of the variable ($x$) was generated and the proposed reconstruction method was applied.

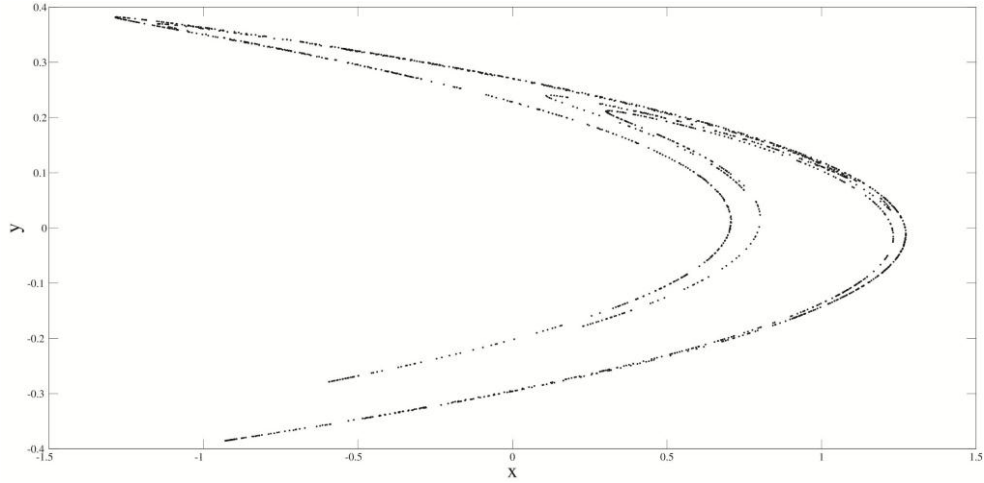

FIG. 6. Strange attractor of the Hénon map.

A brief visual inspection reveals that the Hénon attractor in 2-dimensional space does not contain any persistent holes, which means that its 1-st Betti number is zero. Topological noise exposed through the emergence of short-lived holes is expected. To address this phenomena, we calculated persistent homology for different embedding dimensions. For the embedding dimension 2, the persistent barcode and the persistent diagram for $β_1$ (Fig. 7 left and right, respectively), display emergence of short-lived holes, especially for lower values of the filtering parameter, i.e., the Čech radius. It is easy to notice in the barcode diagram and in the persistent diagram of Fig. 7 that the two points above the diagonal originate from the holes which appear/dissappear between two stripes of the strange attractor.

The same procedure was repeated for the embedding dimension 3. From the persistent barcode (Fig. 8 left) we see that only the 0-th and the 1-st Betti numbers have nonzero values during filtration process. In contrast to the 2-dimensional embedding, 1-holes persist longer, but not enough to escape the topological noise.

Comparing Figures 7 and 8 it is evident that only 1-dimensional holes appear, but they are short-lived and contribute to the topological noise, thus the same phenomena occur even when the embedding dimension is increased. We have checked it for 4-dimensional embedding also, and it follows this same qualitative pattern.

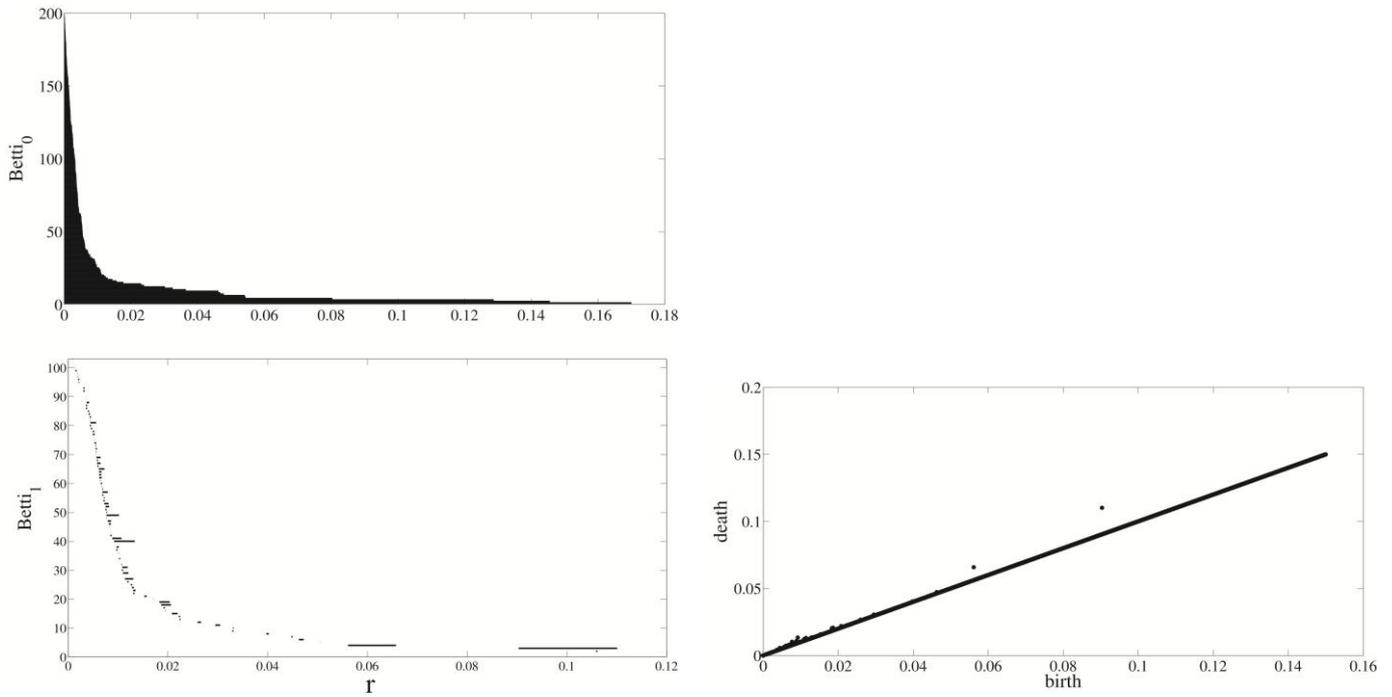

FIG. 7. Persistent barcodes of non-trivial homology groups (left) and persistent diagram (right) of the $H_1$ homology group of the Hénon map for 2-dimensional embedding. To initial and end Čech radii of each barcode on the left diagram correspond the birth and death radii of the diagram on the right.

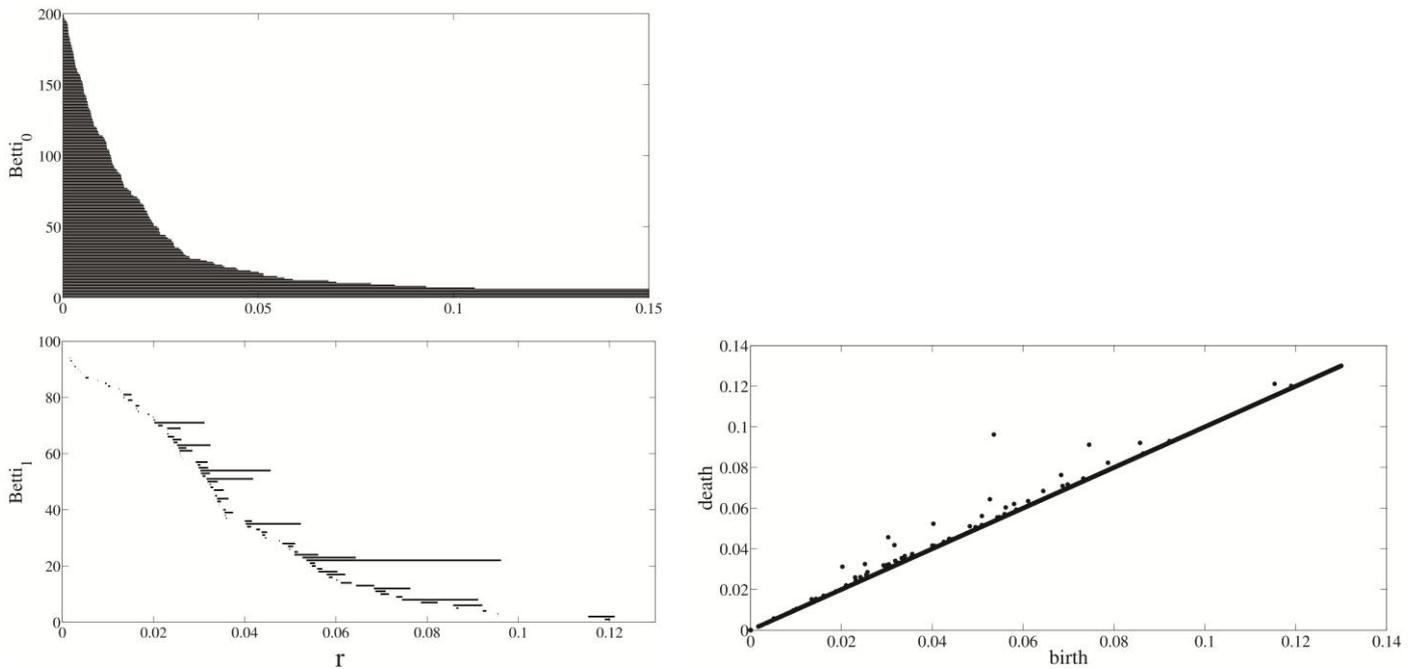

FIG. 8. Persistent barcodes of non-trivial homology groups (left) and persistent diagrams of $H_1$ homology group of the Hénon map for the 3-dimensional embedding.

## B. The Lorenz system

As a paradigm of a higher dimensional systems we consider the Lorenz dynamical system [40] represented by equations:

$$\dot{x} = \sigma(y - x)$$
$$\dot{y} = rx - y - xz$$
$$\dot{z} = xy - bz$$

(2)

where $\sigma$, $r$ and $b$ are parameters. The system is solved using the fourth order Runge-Kutta method for parameter values corresponding to the chaotic regime, namely $\sigma = 10$, $b = 8/3$ and $r = 28$.

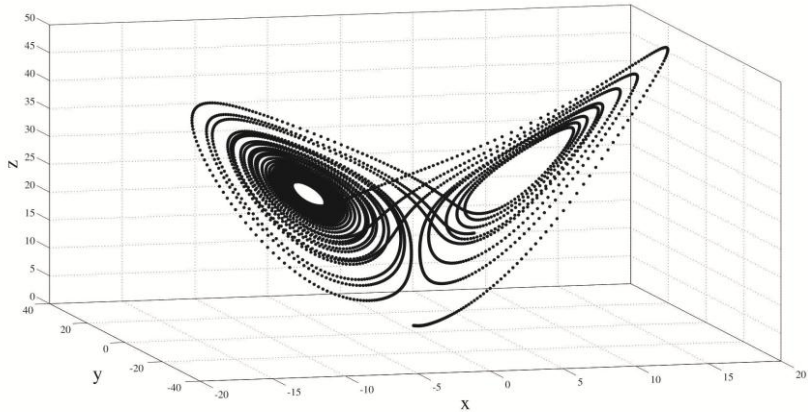

FIG. 9. Strange attractor of Lorenz system.

A time series of the $x$-component is used for the reconstruction procedure and the well-known strange attractor is presented in Fig. 9. A brief visual inspection of the attractor discloses two holes so it would be expected that the persistence barcode and the persistence diagram indicate their appearance. For the embedding dimension equal to 2, the persistence barcode for generators of the 1 − st homology group (Fig. 10 left) clearly indicate persistent lifetime of the two holes. Their lifetime is preserved when the embedding dimension is increased to 3 (Fig. 11 top left) and 4 (Fig. 12 top left).

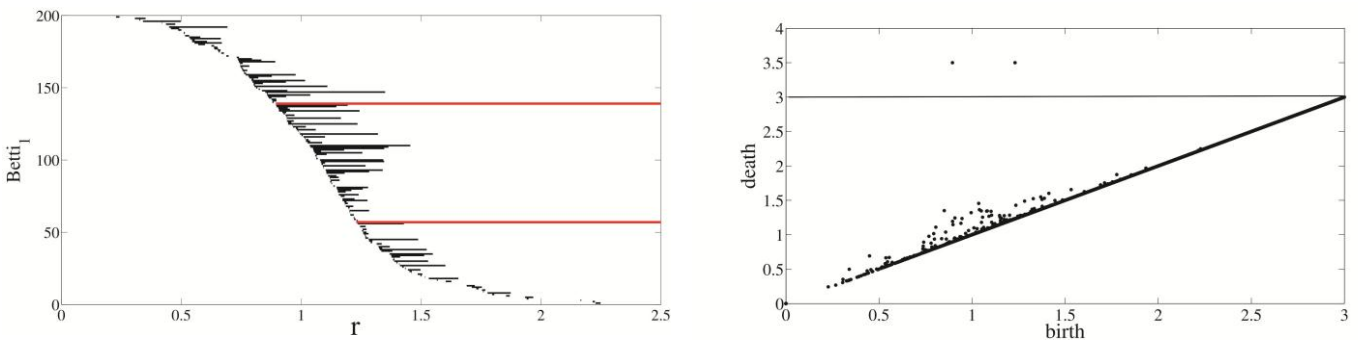

FIG. 10. Persistent barcodes (left) and persistent diagram of the $H_1$ homology group (right) of the Lorenz system for the 2-dimensional embedding.

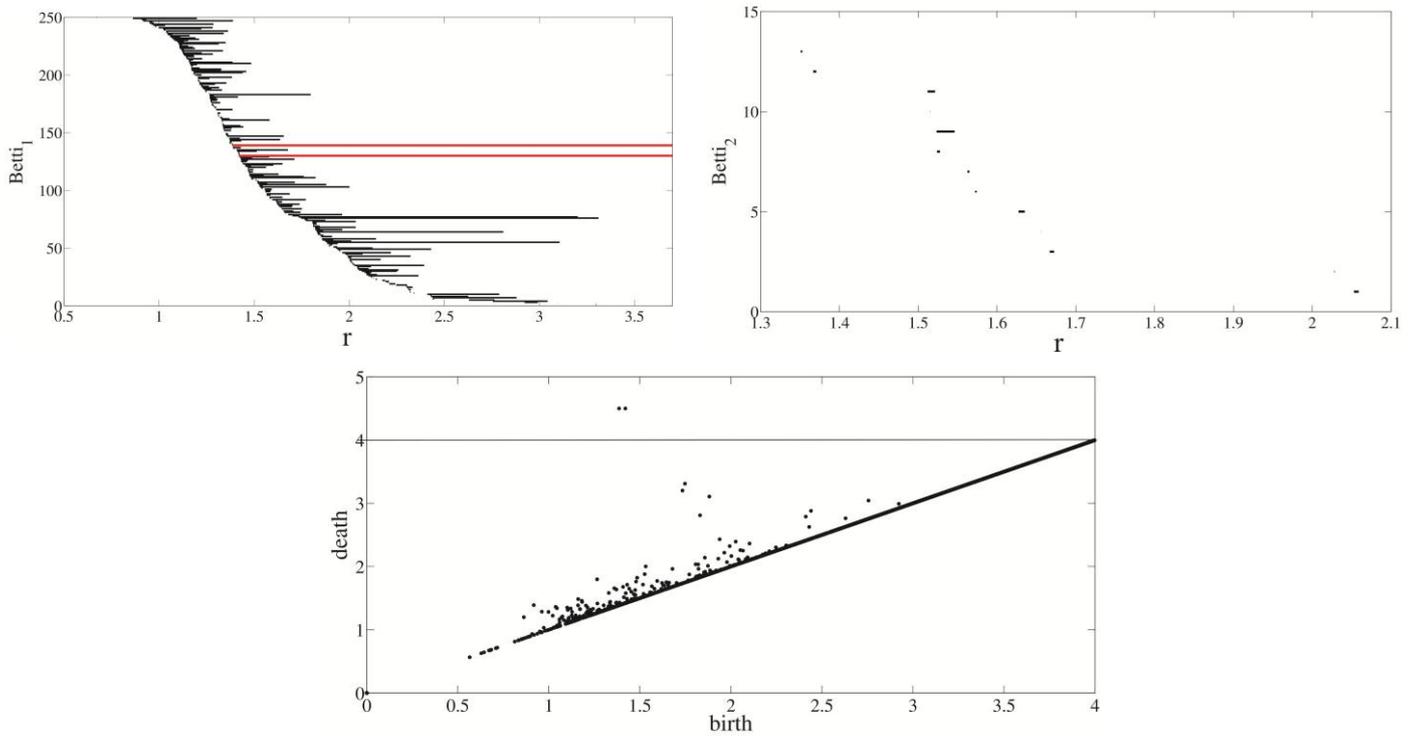

FIG. 11. Persistent barcodes of non-trivial homology groups ($H_1$ and $H_2$) (top) and persistent diagram of the $H_1$ homology group (bottom) of Lorenz system for 3-dimensional embedding.

To gain a better insight into the dynamics of appearance and disappearance of the main topological features of the Lorenz system we call attention to persistent diagrams of the 1−st Betti number for embeddings in dimensions 2, 3, and 4 (Figs. 10 (right), 11 (bottom), and 12 (bottom), respectively). Two points appearing above the horizontal line correspond to two persistent holes that survive filtration. For practical reasons the death-value is chosen arbitrary although its value in terms of persistent homology is actually infinity. With the increase of the embedding dimension the holes which contribute to the topological noise become more distant from the diagonal, but the topological structure is preserved.

An interesting behavior is observed in the cases of 3- and 4-dimensional embeddings, where topological noise for the 2-nd Betti number emerges (11 (top right), and 12 (top right), respectively). A similar behavior is observed in the case of the 5-dimensional embedding suggesting the possibility of persistent topological noise for the 2-nd homology group.

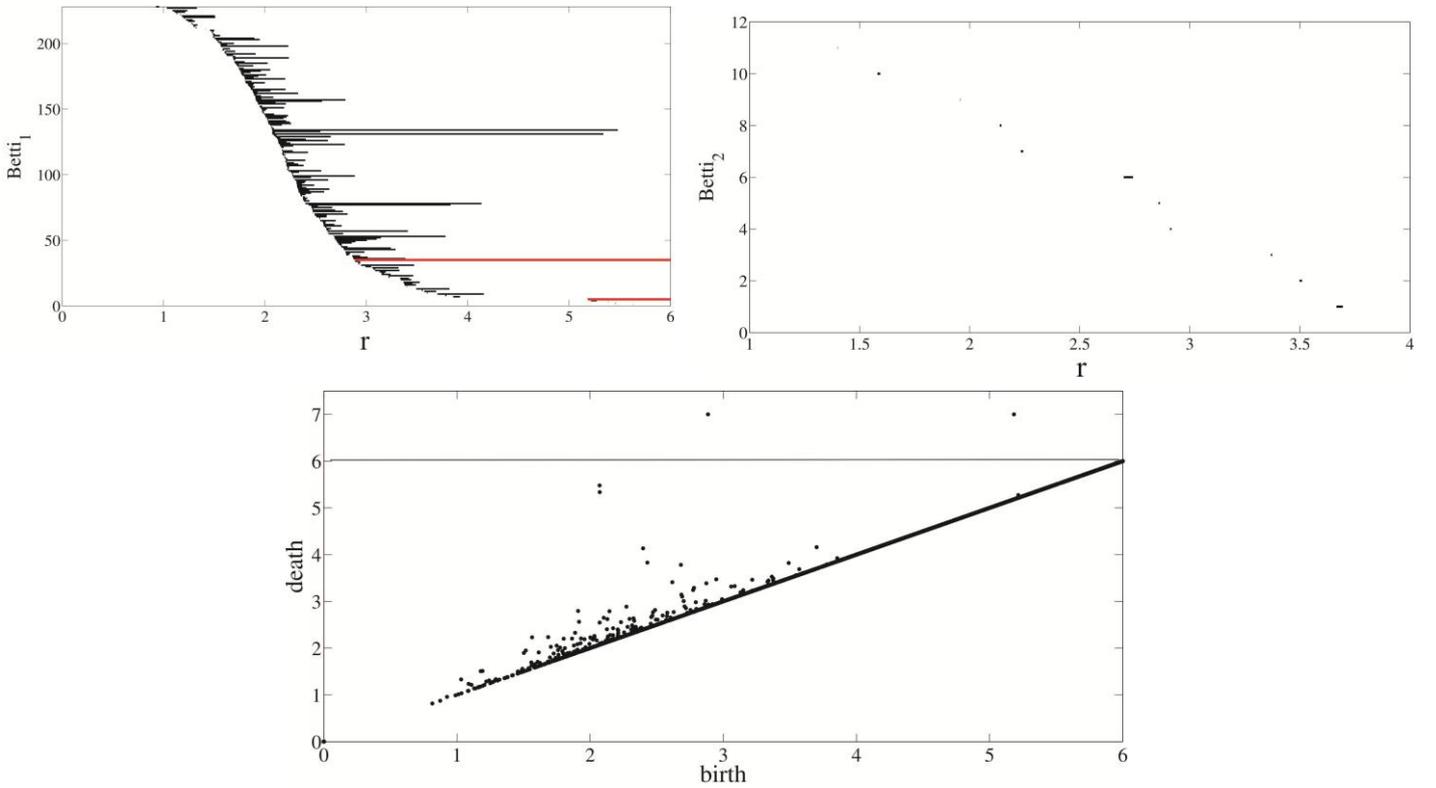

FIG. 12. Persistent barcodes of non-trivial homology groups ($H_1$ and $H_2$) (top) and persistent diagrams of $H_1$ homology group (bottom) of Lorenz system for 4-dimensional embedding.

### C. The Rössler system

So far we have considered systems which are in a particular dynamic regime, specifically in the chaotic state characterized by the existence of a strange attractor. As an example of the phase space reconstruction of a dynamical system in different dynamic regimes we use the Rössler system [41]:

$$\dot{x} = -(y + z)$$
$$\dot{y} = x + ay$$
$$\dot{z} = b + xz - cz$$

(3)

where $a$, $b$ and $c$ are constants. The system is solved using the fourth order Runge-Kutta for different parameter values and we used a time series of the $x$-component for the application of our method. For the same fixed values of $a$ and $b$ (say $a = b = 0.2$) and for different values of constant $c$ the system displays significantly different behavior manifested in the existence of a particular type of the attractor in the phase space. Specifically, we consider the dynamics when the attractor is a period-one limit cycle ($c = 2.3$) Fig. 13 upper left, period-two limit cycle ($c = 3.3$) Fig. 13 upper right, period-three limit cycle ($c = 5.3$) Fig. 13 lower left, and the strange attractor ($c = 6.3$) Fig. 13 lower right.

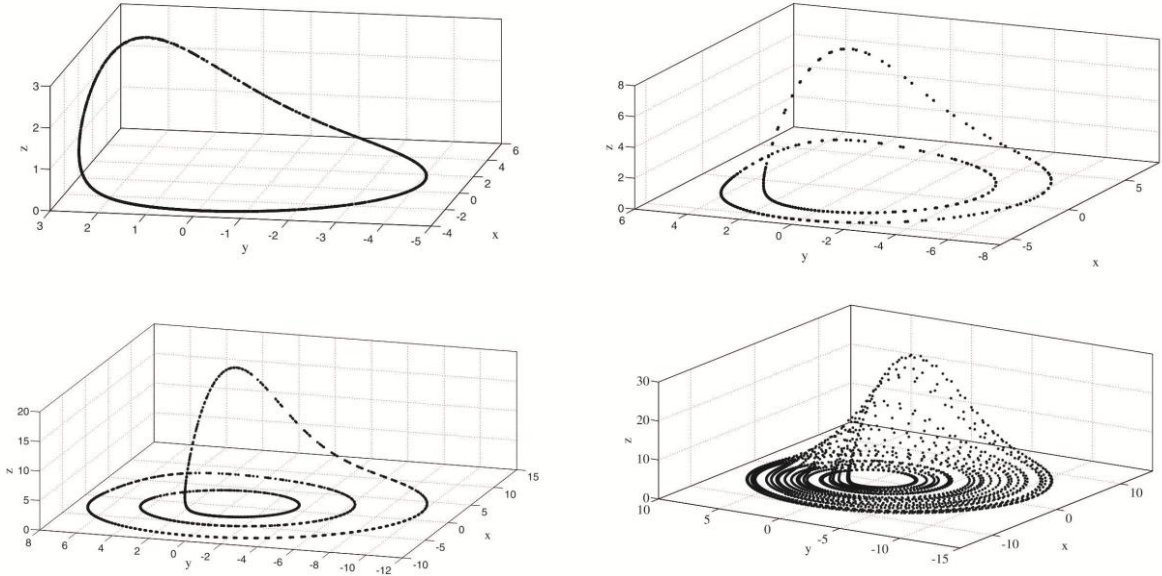

FIG. 13. Attractors for the Rössler system: period-one limit cycle $c = 2.3$ upper left, period-two limit cycle $c = 3.3$ upper right, period-three limit cycle $c = 5.3$ lower left, and strange attractor $c = 6.3$ lower right.

As in the case of the Lorenz system, for embedding dimensions 3 and 4 and for all values of parameter $c$, 2-dimensional topological noise appears due to the emergence of shortlived 2-dimensional holes. Figures 14 and 15 present how the 1-st Betti number changes through filtration for all values of the parameter $c$, and embedding dimensions 2, 3, and 4. The parameter values (Čech radius) for birth and death are properly normalized for easier comparison of the results and points associated to the persistent holes are above the horizontal line. For 2-dimensional embedding (Fig. 14 left) we can see that topological noise does not emerge. On the other hand, for 3- and 4-dimensional embeddings (Fig. 14 right and Fig. 15) it is overwhelmingly present.

Comparison of persistent diagrams corresponding to the non-chaotic and chaotic regimes does not reveal any self-evident disparities. However, difference is noticeable in persistent barcodes of the 1-st and the 2-nd homology group generators in Fig. 16 for the embedding dimension 4 and for parameter values $c = 3.3$ (top left), $c = 5.3$ (top right) and $c = 6.3$ (bottom). For the 4-dimensional embedding we can notice an increased emergence of shortlived 2-dimensional holes, i.e. topological noise.

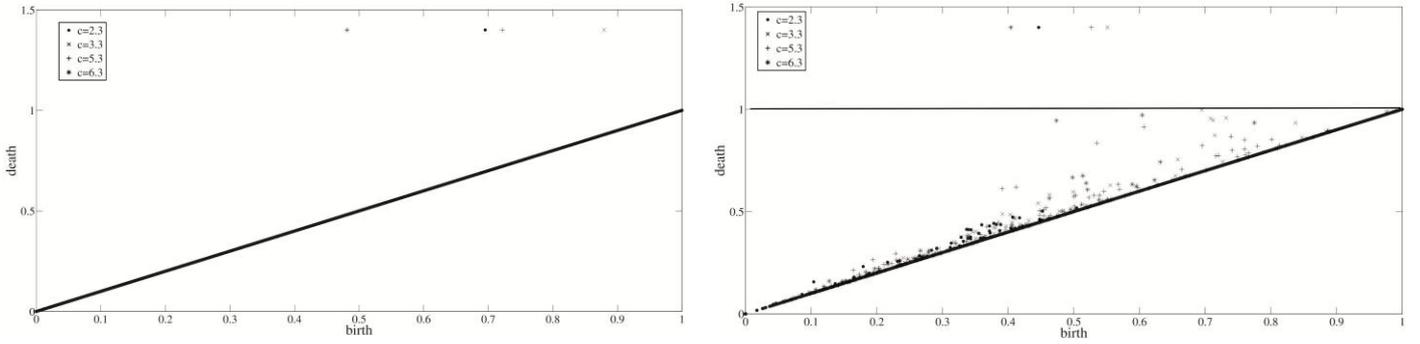

FIG. 14. Persistent diagrams of the $H_1$ homology group of the Rössler system for all considered values of parameter $c$ for the 2-dimensional (left) and the 3-dimensional (right) embedding.

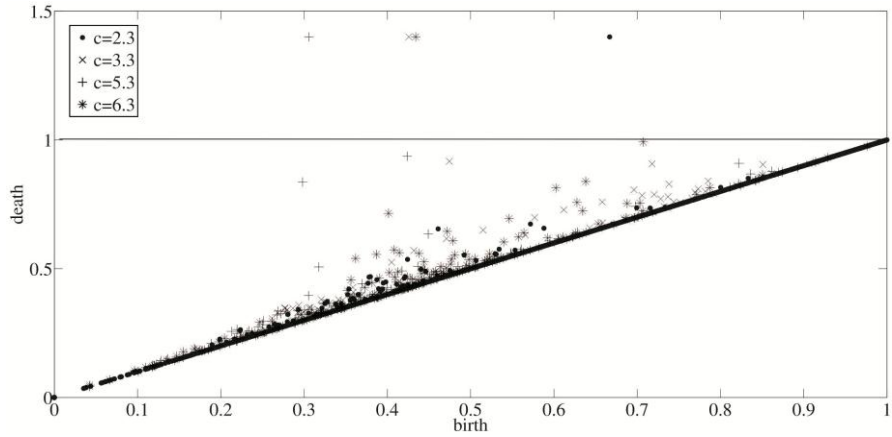

FIG. 15. Persistent diagrams of the $H_1$ homology group of the Rössler system for all consideredvalues of parameter $c$ for the 4-dimensional embedding.

The calculation of the Q-vector is carried out for the clique complex (i.e., simplicial complex built by the maximal cliques) [42] of the 1-skeleton (i.e., of the edges) of the Čech complex at the end of filtration for the embedding dimension 3. In order to compare topological structures embodied in the Q-vectors corresponding to different dynamic regimes the calculated vector values are normalized. The distance between Q-vectors in dynamic regimes corresponding to parameter values $c = 2.3$ and $c = 3.3$ is 1.09. The distance between Q-vectors in regimes of $c = 3.3$ and $c = 5.3$ is 0.88, and between regimes corresponding to parameter values $c = 5.3$ and $c = 6.3$ is 0.54. Thus, as the system moves toward chaotic behavior the distances between Q-vectors decrease. In other words, as the system is closer to the chaotic behavior attractors become more structurally similar.

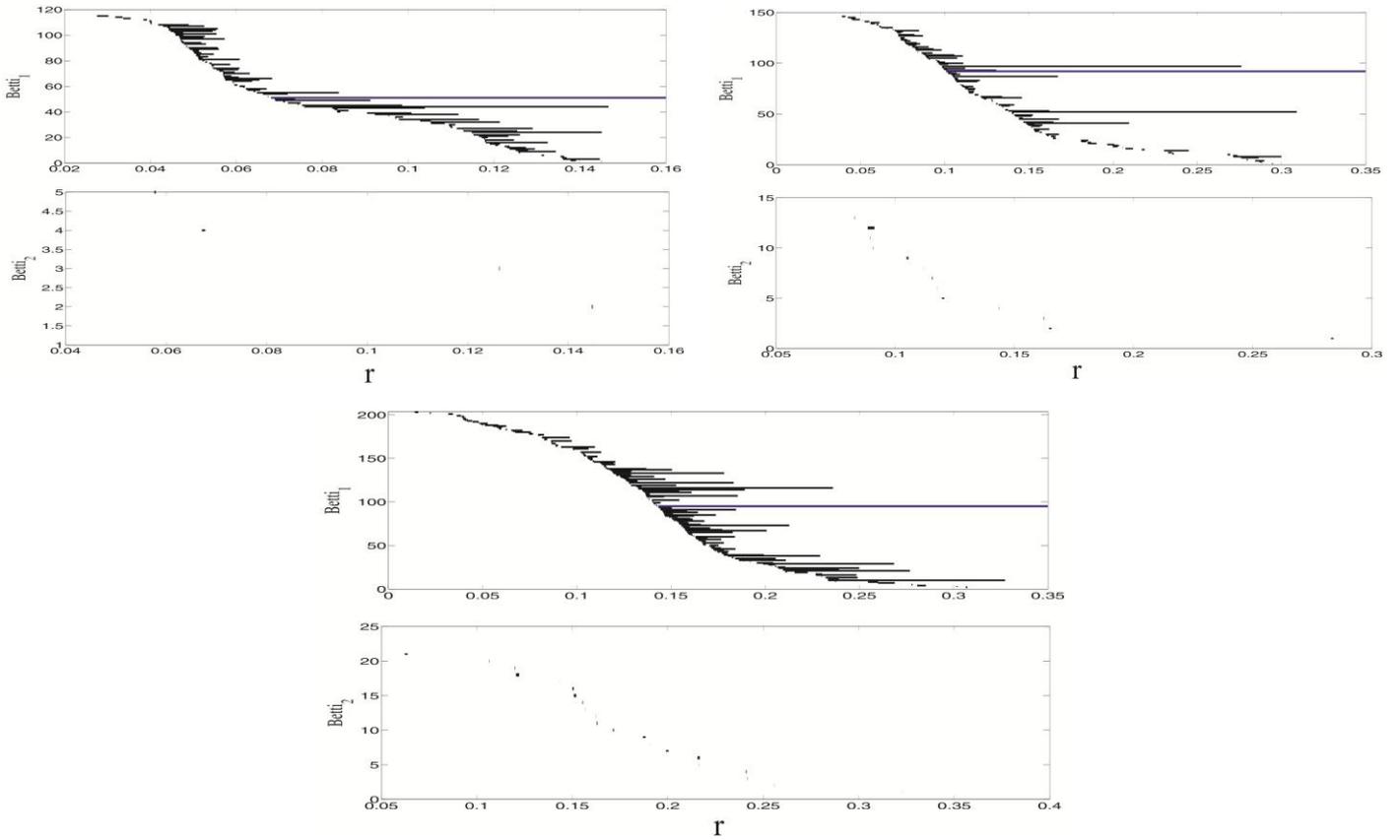

FIG. 16. Barcodes of Rössler system for 4-dimensional embedding for $c = 3.3$ (top left), $c = 5.3$ (top right) and $c = 6.3$ (bottom)

### D. Experimentally obtained data: ECG time series

As a final example of the applications of our method we use an electrocardiograph (ECG) time series measured on the healthy person during sleep. The measured time series is nonlinear (Fig. 17) and it represents a segment taken from a longer ECG recording. Technical details are unimportant for the brief analysis presented here.

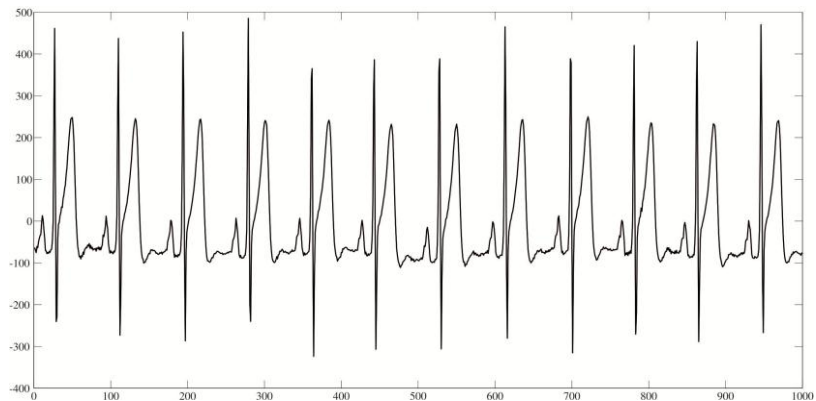

FIG. 17. Time series of the ECG measurement.

A false nearest neighbor method [9] yields an embedding dimension of four. The same methods of topological data analysis are applied as for the numerically generated dynamical systems. The emergence and lifetime of the topological noise is monitored for the embedding dimensions ranging from 3 to 7.

Inspection of the persistent barcode diagrams reveals that topological noise persists at certain dimensions (as for the 2-nd homology group), whereas it vanishes at other dimensions(as for the 3-rd homology group). The results for the 3- dimensional and for the 7-dimensional embedding are presented in Figure 18 left and right, respectively.

Interestingly, topological noise persists at dimensions smaller than the minimal embedding dimension similarly as for dynamical systems in the chaotic regime. As in the cases of chaotic systems, topological noise appears up to a certain dimension of the homology-group and considerably decreases and even disappears at the same dimension when the embedding dimension is increased. This interesting behavior suggests a criterion for the choice of the topologically optimal embedding dimension and points to the significance of the short-lived structures for the topological properties of the nonlinear dynamical systems.

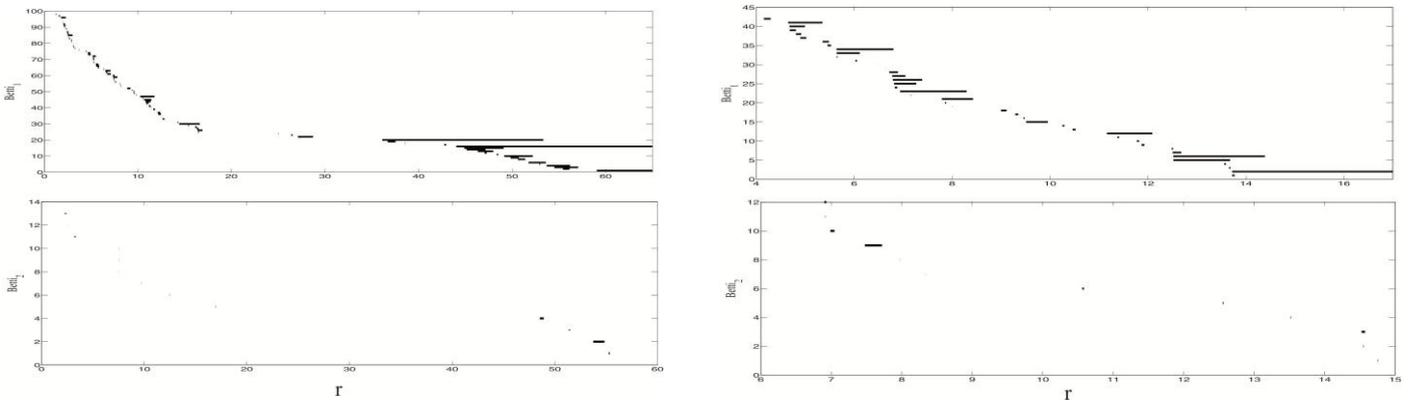

FIG. 18. Barcodes of $H_1$ and $H_2$ homology groups for 3-dimensional (left) and 7-dimensional (right) embedding of ECG.

## IV. SUMMARY AND CONCLUSION

A topological framework combining novel methods of computational topology, such as persistent homology, and standard methods for the analysis of nonlinear dynamic systems is presented. One aspect of this framework represents a general method for constructing a simplicial complex directly from a time series of an observed quantity, which may provide information not attainable from the networks constructed from the same time series. From the aspect of the nonlinear dynamic systems, the method offers new prospects for obtaining information on the topology of the invariant set and parameters necessary for the reconstruction of the state-space topology. One of the important results of the study presented here is that topological noise plays an important role in the delay-coordinate reconstruction procedure and future efforts should be made in order to understand the impact of shortlived topological structures on the reconstruction procedure and the choice of the optimal parameters, namely the embedding dimension and the delay time. In the cases considered topological noise appears at a certain rank of the homology-group and abruptly decreases or even disappears at the same rank when the embedding dimension is increased, suggesting the importance of that particular embedding dimension. A rule for the determination of the topologically relevant embedding dimension may be established based on this property however further studies are needed in order to determine whether this is a generic property. The significance of such behavior points to the importance of topological noise in discovering hidden or short-lived topological structures. The construction of the simplicial complex based on the coarse-graining procedure presented here makes computation more economical and less computationally demanding than some other similar techniques. Solely topological aspect of the difference between the reconstructed and the true attractors puts less stringent requirements on the reconstruction procedure as basic features are observed even at low embedding dimensions, the fact observed also in [30]. This is the consequence of the fact that preservation of topological features in the reconstruction procedure requires only the property of homeomorphism in contrast to the diffeomorphism of the delay-reconstruction procedure. For example, the main macro topological features of the Lorenz system, namely the two holes, are clearly distinguished for the embedding dimension as low as 2. Further applications of new methods of computational topology are envisaged, such as the multidimensional persistent homology [43] as well as applications of the quantities from the Q-analysis [19]. From the aspect of dynamical systems, topological signatures of unstable periodic orbits [44] and topological properties of bifurcations are particularly interesting in relationship with the persistent structures and topological noise.

## ACKNOWLEDGEMENT

The authors S. M. and Y. Z. acknowledge support from the National Nature Science Foundation Committee (NSFC) of China under Project No. 61573119, and a Fundamental Research Project of Shenzhen under Projects No. JCYJ20120613144110654. and No. JCYJ20140417172417109, and M. R. acknowledges financial support by the Serbian Ministry of Education and Science through the project OI 174014.

## REFERENCES

[1] G. Carlsson: Topology and Data, American Mathematical Society. Bulletin. New Series Volume 46, Number 2, 255-08, 2009

[2] N. Packard, J. Crutchfield, D. Farmer, R. Shaw: Geometry from a time series, Phys. Rev. Lett. 45, 712, 1980

[3] S. Maletić and M. Rajković: Combinatorial Laplacian and entropy of simplicial complexes associated with complex networks, Eur. Phys. J. Special Topics 212, 77-97, 2012

[4] M. R. Muldoon, R. S. MacKay, J. P. Huke , D. S. Broomhead: Topology from time series, Physica D 65,

1-16, 1993

[5] F. Takens: Detecting strange attractors in turbulence, in Lecture Notes in Mathematics, No. 898 (Springer-Verlag, 1981)

[6] T. Sauer, J. Yorke, M. Casdagli: Embedology, Journal of Statistical Physics 65, Nos. 3/4, 1991

[7] L.M. Pecora, L. Moniz, J. Nichols, T.L. Carroll: A unified approach to attractor reconstruction, Chaos 17, 013110, 2007

[8] A. M. Fraser, H. L. Swinney: Independent coordinates for strange attractors from mutual information, Phys. Rev. A 33, 1134, 1986

[9] M. B. Kennel, R. Brown, H. D. I. Abarbanel: Determining embedding dimension for phase space reconstruction using a geometrical construction, Phys. Rev. A 45, 3403, 1992

[10] M. E. J. Newman: The structure and function of complex networks. SIAM Rev. 45, 167, 2003

[11] R. V. Donner, M. Small, J. F. Donges, N. Marwan, Y. Zou, R. Xiang, and J. Kurths: Recurrence-based time series analysis by means of complex network methods, Int. J. Bifurcation Chaos 21(04), 1019, 2011

[12] N. Marwan, J. F. Donges, Y. Zou, R. V. Donner, and J. Kurths: Complex network approach for recurrence analysis of time series, Phys. Lett. A 373(46), 4246, 2009

[13] J. Zhang and M. Small: Complex network from pseudoperiodic time series: Topology versus dynamics, Phys. Rev. Lett. 96(23), 238701, 2006

[14] L. Lacasa, B. Luque, F. Ballesteros, J. Luque, and J. C. Nuno: From time series to complex networks: The visibility graph, Proc. Natl. Acad. Sci. U.S.A. 105(13), 4972, 2008

[15] A. H. Shirazi, G. R. Jafari, J. Davoudi, J. Peinke, M. R. Rahimi Tabar andM. Sahim: Mapping stochastic processes onto complex networks, J. Stat. Mech., Theory and Experiment P07046, 2009

[16] X. Sun, M. Small, Y. Zhao, X. Xue: Characterizing system dynamics with a weighted and directed network constructed from time series data, CHAOS 24, 024402, 2014

[17] T. Weng, Y. Zhao, M. Small, D. Huang, Defeng: Time-series analysis of networks: Exploring the structure with random walks, Phys. Rev. E 90, 022804, 2014

[18] Y. Zhao, T. Weng, S. Ye: Geometrical invariability of transformation between a time series and a complex network, Phys. Rev. E 90, 012804, 2014

[19] R. H. Atkin: Combinatorial Connectivities in Social Systems (Birkhäuser Verlag, Base und Stuttgart, 1977)

[20] R. H. Atkin and T. Bastin: A Homological Foundation for Scale Problems in Physics, International Journal of Theoretical Physics 3, 449, 1970

[21] T. Regge: General relativity without coordinates, Nuovo Cimento 19, 558, 1961

[22] A. Bossavit: Computational Electromagnetism : Variational Formulations, Complementarity, Edge

Elements, (Academic Press, 1998)

[23] H. Edelsbrunner and J. L. Harer: Computational topology: An Introduction (American Mathematical Society, Providence, RI, 2010)

[24] S. Maletić and M. Rajković: Consensus formation on simplicial complex of opinions, Physica A 397, 111-120, 2014

[25] S. Maletić, D. Horak and M. Rajković: Cooperation, conflict and higher-order structures of social networks, Advances in Complex Systems, vol. 15, Suppl. No. 1, 1250055, 29, 2012

[26] M. Andjelković, B. Tadić, S. Maletić and M. Rajković: Hierarchical sequencing of online social graphs, Phyisca A 436, 582 -595, 2015

[27] H. Edelsbrunner and J. Harer: Persistent homology - a survey, In Surveys on discrete and computational geometry, volume 453 of Contemp. Math., pages 257-282. Amer. Math. Soc., Providence, RI, 2008

[28] S. Maletić, M. Rajković, and D. Vasiljević: Simplicial complexes of networks and their statistical properties, In Marian Bubak, GeertDick van Albada, Jack Dongarra, and Peter M. A. Sloot, editors, Computational Science ICCS 2008, volume 5102 of Lecture Notes in Computer Science, pages 568-575. Springer Berlin. Heidelberg, 2008

[29] D. Horak, S. Maletić, Milan Rajković: Persistent homology of complex networks, Journal of Statistical Mechanics: Theory and Experiment, P03034, 2009

[30] J. Garland, E. Bradley, J. D. Meiss: Exploring the Topology of Dynamical Reconstructions, arXiv:1506.01128v1

[31] C. M.M. Pereira, R. F. de Mello: Persistent homology for time series and spatial data clustering, Expert Systems with Applications 42, 6026-6038, 2015

[32] V. de Silva and G. Carlsson: Topological estimation using witness complexes. In Symp. On Point-Based Graphics, pages 157-166, 2004

[33] P. Grassberger and I. Procaccia: Measuring the strangeness of strange attractors, Physica D 9, 189-208, 1983

[34] G. Robinson, M. Thiel: Recurrences determine the dynamics, CHAOS 19, 023104, 2009 [35] J. R. Munkres: Elements of Algebraic Topology, (Addison-Wesley Publishing, California, 1984)

[36] R. Ghrist: Barcodes: the persistent topology of data. Bull. Amer. Math. Soc. (N.S.), 45(1), 61-75, 2008

[37] P. Grassberger and I. Procaccia: Characterization of Strange Attractors, Phys.Rev.Lett. 50, 346, 1983

[38] de Silva V and Perry P PLEX home page http://math.stanford.edu/comptoop/programs/plex

[39] M. Hénon: A two-dimensional mapping with a strange attractor, Communications in Mathematical Physics 50 (1): 69-77, 1976

[40] E. N. Lorenz: Deterministic nonperiodic flow, Journal of the Atmospheric Sciences 20 (2), 130-141,

1963

[41] O. Rössler: An Equation for Continuous Chaos, Physics Letters 57A (5), 397-398, 1976

[42] D. Kozlov: Combinatorial Algebraic Topology, (Algorithms and Computation in Mathematics, Springer-Verlag, Berlin Heidelberg, 2008)

[43] E. Carlsson, G. Singh, A. Zomorodian: Computing multidimensional persistence, J. Comp. Geom. 1 (1), 72-100, 2010

[44] K. T. Alligood, T. D. Sauer, J.A. Yorke: Chaos: An Introduction to Dynamical Systems, (Springer-Verlag, New York, 1996)